\documentstyle[12pt]{article}
\begin{document}

\small
\hoffset = -1truecm
\voffset = -2truecm
\title{\bf Gauged Q ball in a piecewise parabolic potential}
\author{
    Xin-zhou Li\footnote {e-mail address: kychz@shtu.edu.cn}\hspace{0.6cm}  Jian-gang Hao\hspace{0.6cm}  Dao-jun Liu\\\footnotesize \it
    Department of physics,Shanghai Normal University, Shanghai 200234 ,China
  \and
   Guang Chen\\\footnotesize \it
    East China Institute for Theoretical Physics, Shanghai 200237,China
     }
\date{}
\maketitle

\begin{abstract}
   Q ball solutions are considered within the theory of a complex scalar field with a gauged
 U(1) symmetry and a parabolic-type potential. In the thin-walled limit, we show explicitly
 that there is a maximum size for these objects because of the repulsive Coulomb force. The
 size of Q ball will increase with the decrease of local minimum of the potential. And when the
 two minima degenerate, the energy stored within the surface of the Q ball becomes significant.
 Furthermore, we find an analytic expression for gauged Q ball, which is beyond the conventional thin-walled limit.
\end{abstract}
\vspace{8cm}
\hspace{1cm}
PACS number(s): 11.27.+d, 11.10.Lm
\newpage

\noindent 1.\hspace{0.4cm}INTRODUCTION

Recently,Theodorakis introduced a piecewise parabolic potential
for a complex scalar field $\phi$ and he showed that it admits
stable Q ball solutions [1]. These solutions have been found
analytically, unlike the case of Polynomial potentials. The Q
ball is a nontopological soliton, and its stability depends on
whether charge can be lost through emission of charged particles
[2,3]. The generation of the Q ball[4] and the possibility of
phase transition precipitated by soliton synthesis[5] have been
actively studied. Later work also examined Q-stars [6] and
D-stars [7,8] with interesting astrophysical implications. The
nontopological soliton can appear in a local U(1) invariant
theory [9-11].

Following Coleman [3], at large distances from the Q ball the
field  should approach the vacuum solution $\phi=0$, if the
charge is to be finite. It turns out that Q balls exist in a
potential $V(\phi)$, for large enough values of the charge Q, if
the function $\frac{2V}{|\phi|^2}$ has a minimum of $\phi$. The
simplest form of potential is a polynomial one where a negative
$|\phi|^4$ term is needed to make potential dip below
$|\phi|^2/2$, and a positive $|\phi|^6$ term to show the
potential bounded from below. However, these potentials give a
difficult task for solving equations of motion. It would be
interesting to obtain analytic solutions of the Q ball.
Fortunately, there is no reason, at least at the classical level,
to restrict ourselves to polynomial potentials. In fact,
Theodorakis [1] has examined a piecewise parabolic potential as
follows

\begin{equation}
U(|\phi|)=\frac{1}{2}\bigg[|\phi|^2+\varepsilon(1-|\phi|)-\varepsilon\bigg|1-|\phi|\bigg|\bigg]
\end{equation}

\noindent In the $1<\varepsilon<2$ case, the global minimum  is
at $|\phi|=0$, but there is a local minimum at
$|\phi|=\varepsilon$, the potential being positive everywhere. In
this paper, we consider the gauged Q balls where the potential is
parabalic-type one. Hitherto, ones were restricted to numerical
methods or thin-walled approximation in the study on gauged Q
ball. In the thin-walled approximation, we find that the radius
of  Q ball is increased with the decline of local minimum of the
potential. And when the two minima are degenerated, the surface
term will play an important role in the fuctional of energy.
Using an iteration method, we have shown an explicit solution of
Q ball with gauge interaction beyond thin-walled limit in this
paper. Furthermore, we can give the asymptotical solutions order
by order in principle.

\vspace{0.8cm}
\noindent 2.\hspace{0.4cm}QUALITATIVE PROPERTIES
OF SOLUTION

We consider the theory with a complex scalar field $\phi$ coupled
to a U(1) gauge field $A_{\mu}$. The Lagrangian density is

\begin{equation}
{\cal L}=\frac{1}{2}\bigg|(\partial_{\mu}- ieA_{\mu})\phi\bigg|^2-U\left(|\phi|\right)- \frac{1}{4} F_{\mu\nu}F^{\mu\nu}
\end{equation}

\noindent where $F_{\mu\nu}=\partial_{\mu}A_{\nu}-\partial_{\nu}A_{\mu}$ and $U(|\phi|)$ is the parabolic-type potential (1).The conserved current and the energy momentum tensor which induce the conserved quantities are

\begin{equation}
j_{\mu}=\frac{i}{2}\bigg[\phi(\partial_{\mu}+ieA_{\mu})\bar{\phi}-\bar{\phi}(\partial_{\mu}-ieA_{\mu})\phi\bigg]
\end{equation}

\noindent and

\begin{equation}
T_{\mu\nu}=\frac{1}{2}\bigg[(\partial_{\mu}-ieA_{\mu})\phi(\partial_{\nu}+ieA_{\nu})\bar{\phi}+(\partial_{\nu}-ieA_{\nu})\phi(\partial_{\mu}+ieA_{\mu})\bar{\phi}\bigg]-F_{\mu\sigma}F^{\sigma}_{\nu}-{\cal L}\eta_{\mu\nu}
\end{equation}

\noindent respectively. To find the solutions of Q ball, we use
$ans\ddot{a}tz$ as follows

\begin{equation}\phi=f(r)e^{i\omega t}\end{equation}

\noindent and

\begin{equation} A_{\mu}=A_0(r)\delta_{\mu0}\end{equation}

\noindent where we assume $\omega>0$ for definiteness. The lowest
energy state will have no electric currents and therefore no
magnetic fields. The spatial components of gauge potential are
zero as there is no magnetic field. We choose a boundary
condition $A_{0}\rightarrow 0$ as $r\to\infty$.

Using the above configurations, the Lagrangian becomes

\begin{equation}
L=4\pi\int drr^2\bigg[-\frac{1}{2}f'^2+\frac{1}{2e^2}g'^2+\frac{1}{2}g^2f^2-U(f)\bigg]
\end{equation}

\noindent where $g=\omega-eA_{0}(r)$, and prime denotes $d/ dr$.
The Noether charge associated with the U(1) symmetry becomes

\begin{equation}
Q=\int d^3xgf^2
\end{equation}

\noindent When we insert Eqs. (5) and (6) in the general field
equation deduced from Eq. (2) we find the following equations of
motion:

\begin{equation}
f''+\frac{2}{r}f'+fg^2=f-\varepsilon \hspace{0.1cm}, \hspace{1.2cm} r\le R, f\ge 1
\end{equation}

\begin{equation}
f''+\frac{2}{r}f'+fg^2=f \hspace{0.1cm},  \hspace{2cm} r>R, f<1
\end{equation}

\begin{equation}
g''+\frac{2}{r}g'=e^2f^2g \hspace{0.1cm}. \hspace{4.5cm}
\end{equation}

\noindent The fact that the field equations (9-11) are nonlinear
implies that the analytic solutions that will be found will be
approximate. To gain insight into the Q ball for $e\not=0$, we
show some qualitative properties of the solution. From (8) and
(11) we have

\begin{equation}
g(r)\to \omega-\frac{e^2Q}{4\pi r}
\end{equation}

\noindent as $r \to \infty$. Then Eq. (10) can be reduced to the
asymptotic form

\begin{equation}
\frac{1}{r}(rf)''+(\omega^2-1)f=0
\end{equation}

For the approximate equation (13), we have a solution as follows

\begin{equation}
f(r)=f_0 \exp(-\sqrt{1-\omega^2}r)/r
\end{equation}

\noindent Clearly, a necessary condition for the existence of a
solution is $\omega < 1$. Additionally, for the solution to be
well-behaved at the origin, $f'$ and $g'$ should approach zero at
least faster than $r$ for $r\to 0$. By using the boundary
conditions of $f$ and $g$ the energy functional can be written as

\begin{equation}
E=\frac{1}{2}\omega Q+4\pi\int r^2dr\big[\frac{1}{2}f'^2+V(f)\big]
\end{equation}

\noindent For the $e\not=0$ case, we expect that the energy will
be  increased over the $e=0$ case due to Coulomb repulsion, with
Coulomb energy becoming more important as Q becomes large.
Therefore, if $\partial E/\partial Q>1$, we must consider that
some charge can be put into the interior region of the Q ball and
some can be put in free particles. As discussed above, there
exists a $Q_{min}$ and a $Q_{max}$, so that when $Q>Q_{min}$ the
soliton is quantum mechanically stable and when $Q>Q_{max}$ the
lowest energy state of the system is composed of a soliton with
charge $Q_{max}$ together with free particles carrying charge
$Q-Q_{max}$. In the $e=0$ theory, since $Q_{max}\to \infty$, the
condition $Q_{min}<Q_{max}$ is always satisfied.

\vspace{0.8cm}
 \noindent 3.\hspace{0.4cm}THIN-WALLED APPROXIMATION

 If we choose a continuous $ans\ddot{a}tz$ for the scalar field,
 the most convenient one might be the piecewise $ans\ddot{a}tz$

 \begin{equation}
  f(r)= \left\{
  \begin{array}{r@{\quad \quad}c}
   \tilde{f} & r < R  \\
  \tilde{f}- \frac{\tilde{f}}{\delta}(r-R) & R\leq r \leq R + \delta \\
  0 & r>R+ \delta
  \end{array} \right.
 \end{equation}

 \noindent where $\delta$ is the width of the wall and $\tilde{f}$
 is a constant. But unfortunately, this $ans\ddot{a}tz$,unlike $e=0$ case, does not
 satisfy the equations of motion of the scalar field. Therefore,
 one must choose the thin-walled approximation in which the
 spatial configuration of the scalar field $f(r)$ can be
 considered as a step function.

In the thin-walled approximation, the spatial configuration of the scalar
 field $f(r)$ can be considered as a step function

\begin{equation}
f(r)=\tilde{f}\theta(R-r)
\end{equation}

\noindent  Although the gradient of $f$ in the thin surface is a
delta function, and hence the associated density of kinetic
energy is infinite, the surface term in the integral of energy is
finite. Especially,in the case of $\varepsilon<2$, the
contribution of the kinetic energy term to the integral of energy
is negligible. Therefore the constant $\tilde{f}$ in Eq. (17) is
determined by minimizing energy functional. In this case, we can
solve Eq. (11)

\begin{equation}
g(r)=\frac{(\omega-\frac{e^2Q}{4\pi R})R}{\sinh(e\tilde{f}R)}\frac{\sinh(e\tilde{f}r)}{r} \hspace{1.5cm}for\hspace{0.3cm} r\le R
\end{equation}

\begin{equation}
g(r)=\omega-\frac{e^2Q}{4\pi r}\hspace{3.8cm}for\hspace{0.3cm} r>R
\end{equation}

\noindent where the gauge is chosen so that $A_0\to 0$ for $r\to
\infty$. Inserting the above solution into Eq.(8), we have

\begin{equation}
\omega=\frac{e^2Q}{4\pi R}\bigg[1-\frac{\tanh(e\tilde{f}R)}{e\tilde{f}R}\bigg]^{-1}
\end{equation}

The total energy of the ball is reduced to

\begin{equation}
E=\frac{e^2Q^2}{8\pi
R}\bigg[1-\frac{\tanh(e\tilde{f}R)}{e\tilde{f}R}\bigg]^{-1}+\frac{4\pi}{3}U(\tilde{f})R^3
\end{equation}

\noindent From $\partial E/\partial R\bigg|_{R=R_{\ast}}=0$, the
radius $R\ast$ of the ball is determined by

\begin{equation}
4\pi\sqrt{2U(\tilde{f})/\tilde{f}^2}R\bigg[\frac{e\tilde{f}R}{\tanh(e\tilde{f}R)}-1\bigg]=e^2Q
\end{equation}

\noindent It is a formidable task to find exact solutions of the
transcendental equation (22). We have to use $e\tilde{f}R\ll 1$
approximation [9]. Eq.(21) can be reduced to the following form

\begin{equation}
E=Q\bigg[\frac{2U(\tilde{f})}{\tilde{f}^2}\bigg]^{\frac{1}{2}}\bigg[1+\frac{(\frac{3}{4\pi} \tilde{f}Q)^{\frac{2}{3}}(\tilde{f}^2/2U(\tilde{f}))^{\frac{1}{3}}}{5}\bigg]
\end{equation}

\noindent Minimizing the energy with respect to $\tilde{f}$ and
$Q(R)$ we find that

\begin{equation}
\tilde{f}_{\ast}=2-\frac{8e^2}{5\varepsilon}\bigg[\frac{Q^2(1-\frac{\varepsilon}{2})^2}{3\pi}\bigg]^{\frac{1}{3}}
\end{equation}

\begin{equation}
R_{\ast}=\bigg[\frac{3Q}{8\pi\sqrt{2(2-\varepsilon)}}\bigg]^{\frac{1}{3}}\bigg[1+\frac{1}{45}\bigg(\frac{3e^3Q}{\pi\sqrt{2(2-\varepsilon)}}\bigg)^{\frac{2}{3}}\bigg]
\end{equation}

\noindent and

\begin{equation}
E_{\ast}=Q\sqrt{1-\frac{\varepsilon}{2}}\bigg\{1+\frac{1}{5}\bigg(\frac{3e^3Q}{\pi\sqrt{2(2-\varepsilon)}}\bigg)^{\frac{2}{3}}\bigg\}
\end{equation}

\noindent For a fixed charge $Q$, we see that both the radius and
the energy are larger than ungauged one. Furthermore, from the
condition $\partial E/\partial Q\bigg|_{Q=Q_{max}}=1$, one can
obtain the upper bound on the total charge $Q_{max}$ as

\begin{equation}
Q_{max}=\frac{2\pi}{3e^2}\bigg[5(\sqrt{\frac{2}{2-\varepsilon}}-1)\bigg]^{\frac{3}{2}}\sqrt{\frac{2-\varepsilon}{2}}
\end{equation}

\noindent When $Q>Q_{max}$ the lowest energy state of the system
is composed of a Q ball with charge $Q_{max}$ together with free
particles carrying charge $Q-Q_{max}$. Here note that the
inconsistency of the approximation $e\tilde{f}R\ll1$ when
$\varepsilon$ approaches 2. In the $\varepsilon=2$ case,we must
considered the surface term with a finite width. If
$e\tilde{f}R\gg1$, as will be discussed later, the energy can be
approximately written by

\begin{equation}
E=\frac{e^2Q^2}{8\pi R}+ 4\pi\varepsilon R^2
\end{equation}

\noindent From $\partial E/\partial R\bigg|_{R=R_{\ast}}=0$, the
radius of the Q ball is given by

\begin{equation}
R_{\ast}=\bigg(\frac{e^2Q^2}{128\pi^2}\bigg)^{1\over 3}
\end{equation}

\noindent and

\begin{equation}
E_{\ast}=E(R_{\ast})=\frac{3(eQ)^{4\over 3}}{2^{2\over 3}\pi^{1\over 3}}
\end{equation}

\noindent The upper bound on charge $Q_{max}$ is also given by

\begin{equation}
Q_{max}=\frac{\pi}{2e^4}
\end{equation}

\noindent For simplicity, we write $f_{\ast}$, $R_{\ast}$ and
$E_{\ast}$ as $f$, $R$ and $E$ hereafter.

\vspace{0.8cm} \noindent 4.\hspace{0.4cm}BEYOND THIN-WALLED LIMIT

We may use the successive approximation method of differential
equation for higher order approximation of the Q ball solution.
Using $rg(r)=e^{u(r)}$ and $v(r)=u'(r)$, Eq.(11) can be reduced to

\begin{equation}
v'=e^2f^2-v^2
\end{equation}

\noindent with the initial condition $v=v_0$ for $r=r_0$. It can
be written in the form

\begin{equation}
v=v_0+\int^r_{r_0}(e^2f^2-v^2)dr
\end{equation}

\noindent Substituting the thin-walled solution $v_1(r)$ instead
of $v$ to the right member, we obtain a new function $v_2$
different $v_1$, unless $v_1$ is a exact solution of Eq.(32).
Substituting $v_2$ instead of $v$ to the right member of Eq. (32),
we obtain a function $v_3$ and so on. The sequence $v_1$, $v_2$,
$\cdots$, $v_n$, $\cdots$ obtained in this way is convergent in a
certain interval containing $r_0$ to the desired solution of the
given equation, provided that the assumptions of Cauchy's
existence theorem. In principle, we can find the exact solution
$g(r)$ if $f(r)$ has been known. This method is also called  the
iteration method.

In the weak coupling situation $e\ll1$ and $g(r)=\omega$, we have
solutions of Eq. (9) and (10) as follows

\begin{equation}
f=a+b\frac{\sinh(\nu r)}{r} \hspace{3cm} r\le R
\end{equation}

\begin{equation}
f=\frac{R}{r}e^{\nu(R-r)} \hspace{3.6cm} r>R
\end{equation}

\noindent where $\nu =\sqrt{1-\omega^2}$,
$a=\frac{\varepsilon}{\nu^2}$ and
$b=(1-{\varepsilon\over{\nu^2}})\frac{R}{\sinh \nu R}$. For
thin-walled limit, we have $\tilde{f}=a+b\nu$ in the Eq. (19).
Equivalently, we have

\begin{equation}
v_1(r)=e(a+b\nu)\coth[e(a+b\nu)r] \hspace{3cm} r\le R
\end{equation}

\begin{equation}
v_1(r)=\bigg(r-\frac{e^2Q}{4\pi\omega}\bigg)^{-1}\hspace{4.6cm} r>R
\end{equation}

\noindent Substituting $v_1(r)$ instead of $v$ to Eq. (33), we
obtain high-order approximation

\begin{eqnarray}
v_2&=&e^2a^2r+2e^2ab Shi(\nu r)+e^2b^2\nu Shi(2\nu r)-\frac{e^2b^2}{2}\frac{\sinh^2(\nu r)}{r}\nonumber \\
& &+e(a+b\nu)\{\coth[e(a+b\nu)r]-e(a+b\nu)r\}+C \hspace{3cm} r\le R\nonumber\\
v_2&=&\bigg(r-\frac{e^2Q}{4\pi \omega}\bigg)^{-1}-\frac{e^2c^2e^{-2\nu r}}{r}-2e^2R^2\nu\exp(2\nu R)Ei(-2\nu r)\hspace{1.3cm} r>R
\end{eqnarray}

\noindent where $Shi(x)$ is the hyperbolic-sine-integral function
and $Ei(x)$ the exponential-integral function. Using the relation

\begin{equation}
Ei(-2\nu r)=-e^{2\nu r}\int^{\infty}_1\frac{1}{2\nu r+\ln t}\frac{dt}{t^2}
\end{equation}

\noindent one can easily find that $v(r)$ satisfies the boundary
condition at space infinity $v(\infty)=0$. The integral constant
$C$ can be fixed by continuity at $r=R$. Therefore, we have

\begin{eqnarray}
g_2&=&\bigg(\omega-\frac{e^2Q}{4\pi R}\bigg)\frac{\sinh[e(a+b\nu)r]}{\sinh[e(a+b\nu)R]}\bigg(\frac{r}{R}\bigg)^{\frac{e^2b^2}{2}-1}\exp\{\frac{e^2}{2}(2ab\nu+b^2\nu^2)(R^2-r^2)\hspace{3cm}\nonumber\\
& &+2e^2ab[rShi(\nu r)-RShi(\nu R)]+\frac{2}{\nu}e^2ab[\cosh(\nu R)-\cosh(\nu r)]\hspace{3cm}\nonumber\\
& &+e^2b^2\nu[rShi(2\nu r)-RShi(2\nu R)]+e^2b^2[\cosh(2\nu R)-\cosh(2\nu r)]\hspace{3cm}\nonumber\\
& &+e^2R^2\nu +2e^2R^2\nu \exp(2\nu R)\int^\infty_R Ei(-2\nu r)dr \}\hspace{5cm} r\le R
\end{eqnarray}

\begin{eqnarray}
g_2&=&\bigg(\omega-\frac{e^2Q}{4\pi R}\bigg)\exp\{e^2R^2\nu \exp(2\nu R)[Ei(-2\nu r)+\int^\infty_r Ei(-2\nu r)dr]\}\hspace{2cm} r>R\hspace{1.5cm}
\end{eqnarray}

\noindent Eq. (34), (35), (40) and (41) are the solution of
high-order approximation beyond the thin-walled limit. Our
approximate analytic solution is well in agreement with the
numerical results [9], which have been obtained by many authors.
The study on Q ball in a parabolic potential, as a toy model, can
enable us to understand the Q ball analytically, which provide
more information than numerical study. In principle, by
continuing the procedures mentioned above, we can obtain the
approximate solution with any high-order accuracy. It's obvious
that this method can also be employed in investigating other
spherically symmetric soliton, and the corresponding results will
appear elsewhere.

\vspace{10cm}\hspace{2cm} \special{bmp:d:/fig1.bmp x=10cm y=10cm}

\vspace{0.8cm}
\noindent ACKNOWLEDGMENTS

This work was partially supported by National Nature Science
Foundation of China under Grant No. 19875016, and National Doctor
Foundation of China under Grant No. 1999025110.


\newpage
\begin{thebibliography}{99}
\bibitem {Theodorakis} S. Theodorakis, Phys. Rev. {\bf D61}, 047701 (2000).
\bibitem {Lee} T. D. Lee and Y. Pang, Phys. Rep. {\bf 221}, 251 (1992).
\bibitem {Coleman} S. Coleman, Nucl. Phys. {\bf 262}, 263 (1985).
\bibitem {Griest} K. Griest and E. W. Kolb, Phys. Rev. {\bf D40}, 3231 (1989);
J. A. Frieman, A. V. Olinto, M. Gleiser, and C. Alcock, Phys. Rev. {\bf D40}, 3241 (1989);
K.Griest, E. W. Klob, and A. Massarotti, Phys. Rev. {\bf D40}, 3529 (1989).
\bibitem {kusenko} A. Kusenko, Phys. Lett. {\bf B405}, 108 (1997); {\bf B406}, 26 (1997);
A. kusenko, M. Shaposhnikov, Phys. Lett. {\bf B418}, 46 (1998);
A. Kusenko, M. Shaposhnikov, P. G. Tinyakov, and I. Tkachev, Phys. Lett. {\bf B423}, 104 (1998);
G. Dvali, A. Kusenko, and M. Shaposhnikov, Phys. Lett. {\bf B417}, 99 (1998).
\bibitem {Lynn} B. W. Lynn, Nucl. Phys. {\bf B321}, 465 (1989).
\bibitem {Li} X. Z. Li and X. H. Zhai, Phys. Lett. {\bf B364}, 212 (1995).
\bibitem {Li1} X. Z. Li, X. H. Zhai and G. Chen, Astropart. Phys. {\bf 13}, 245 (2000).
\bibitem {Lee1} K. Lee, J. A. Stein-Schabes, R. Watkins and L. M. Widrow, Phys. Rev. {\bf D39}, 1665 (1989).
\bibitem {Shi} X. Shi and X. Z. Li, J. Phys. {\bf A24}, 4075 (1991).
\bibitem {Li2} X. Z. Li, Z. Ni and J. Zhang, J. Phys. {\bf A27}, 507 (1994).

\end{thebibliography}
\end{document}